\documentclass[preprint,aps,prd,groupedaddress,floatfix,
nofootinbib]{revtex4}
\usepackage{graphicx}
\usepackage{epsfig}
\usepackage{graphicx}
\usepackage{longtable}

\usepackage{hyperref}

\usepackage{amsmath,amssymb,bm}

\newcommand{\mathd}{\mathrm{d}}

\def\Comment#1{}
\newcommand{\bean}{\begin{eqnarray*}}
\newcommand{\eean}{\end{eqnarray*}}

\newcommand{\gapproxeq}{\lower
.7ex\hbox{$\;\stackrel{\textstyle >}{\sim}\;$}}
\newcommand{\lapproxeq}{\lower
.7ex\hbox{$\;\stackrel{\textstyle <}{\sim}\;$}}

\newcommand\lsim{\mathrel{\rlap{\lower4pt\hbox{\hskip1pt$\sim$}}
    \raise1pt\hbox{$<$}}}
\newcommand\gsim{\mathrel{\rlap{\lower4pt\hbox{\hskip1pt$\sim$}}
    \raise1pt\hbox{$>$}}}
\newcommand{\ba}{\begin{array}}
\newcommand{\ea}{\end{array}}
\newcommand{\nn}{\nonumber}

\newcommand{\be}{\begin{equation}}
\newcommand{\ee}{\end{equation}}
\newcommand{\bear}{\begin{eqnarray}}
\newcommand{\eear}{\end{eqnarray}}

\newcommand{\ket}{\,\rangle}
\newcommand{\bra}{\langle \,}
\def\bat{\begin{array}{cc}}

\begin{document}
\preprint{\vbox{\hbox{BARI-TH-2012-659 \hfill}}}
\title{Holography and chiral Lagrangian}
\author{Fen Zuo\footnote{Email: \textsf{fen.zuo@ba.infn.it}}}
\affiliation{Istituto Nazionale di Fisica Nucleare, Sezione di Bari, Italy}

\begin{abstract}
In a wide class of holographic models described by the Yang-Mills and Chern-Simons terms,
we derive all the ${\cal O}(p^6)$ Chiral Perturbation Theory low-energy constants.
Various model-independent relations exist among
the constants up to this order, generalizing the previously found relations at $p^4$ order.
Some of them can be extended to relations between different amplitudes.
\\
\end{abstract}

\maketitle

\section{Holographic setup}
In the $N_c\to \infty$ limit, one can introduce infinite number of resonances into the nonlinear sigma models, in the spirit of hidden local symmetry. The resulting theory describes a gauge field propagating in a curved 5-dimensional spacetime with two boundaries, where the gauge symmetries are
identified as the global chiral symmetries. The theory is described by the 5D Yang-Mills plus Chern-Simons action~\cite{Son:2003et}
 \begin{eqnarray}
  S_{\rm YM} &=& -\int d^5x ~{\rm tr} \left[-f^2(z){\cal F}_{z\mu}^2
  + \frac{1}{2g^2(z)}{\cal F}_{\mu\nu}^2 \right], \nonumber\\
  S_{\rm CS} &=& -\frac{N_c}{24\pi^2} \int\! {\rm tr}
  \left[{\cal AF}^2+\frac{i}{2}{\cal A}^3{\cal F}-\frac{1}{10}{\cal A}^5
\right].\nonumber
\end{eqnarray}
The string construction of this kind of models was later obtained~\cite{Sakai:2004cn} by introducing the $D8/\overline{D8}$ flavor branes into the $D4$ background, with the above action given by the low energy effective action of the flavor branes. Spontaneous chiral symmetry breaking was implemented geometrically as the smoothly connection of the flavor branes and anti-branes in the infrared.

Since the theory originates from the nonlinear sigma model, it implements automatically a massless Goldstone boson, which can be compactly described through
the Wilson line
\begin{equation}
U(x)=\mbox{P} \exp\left\{i\int^{+z_0}_{-z_0} {\cal A}_z(x,z') dz'\right\}.
\end{equation}
At the meantime, the gauge bosons of the infinitely many hidden local symmetries can be interpreted as the vector and axial mesons, which are
included in the gauge potential ${\cal A}_\mu$. External sources can be introduces by gauging ${\cal A}_\mu$ at the two boundaries, which is then solved by
\begin{equation}
{\cal A}_\mu(x,z)=\ell_\mu(x) \psi_-(z)+r_\mu(x) \psi_+(z)+\sum_{n=1}^\infty v_\mu^n(x)\psi_{2n-1}(z)+\sum_{n=1}^\infty a_\mu^n(x)\psi_{2n}(z),
\end{equation}
with $\psi_\pm=(1\pm\psi_0)/2$.

\section{$\chi$PT Lagrangian up to ${\cal O}(p^6)$}
The action in the previous section is constructed in the gauge ${\cal A}_M\to 0$ ($z\to \pm z_0$). One can choose a suitable gauge transformation to go to another gauge ${\cal A}_z=0$, in which one finds:
\begin{equation}
{\cal A}_\mu(x,z)=i\Gamma_\mu(x)+\frac{u_\mu(x)}{2}\psi_0(z)+\sum_{n=1}^\infty v_\mu^n(x)\psi_{2n-1}(z)+\sum_{n=1}^\infty a_\mu^n(x)\psi_{2n}(z),
\end{equation}
with $\Gamma_\mu$ and $u_\mu$ the commonly used 4D operators in chiral perturbation theory~($\chi$PT).
one then obtains a chiral lagrangian with infinitely many resonances by substituting this into the gauge-transformed action. The terms without resonances give the $\chi$PT lagrangian up to ${\cal O}(p^4)$, with the expressions for the low-energy constants~(LECs)~\cite{Hirn:2005nr,Sakai:2005yt} \begin{eqnarray}
&&f_\pi^2=4\left(\int_{-z_0}^{z_0}\frac{\mathd z}{f^2(z)}\right)^{-1},~~~L_1=\frac{1}{2} L_2=-\frac{1}{6}L_3=\frac{1}{32}\int_{-z_0}^{z_0} \frac{(1-\psi_0^2)^2}{g^2(z)}  \, \mathd z \, ,
\nonumber\\
&&L_9=-L_{10}=\frac{1}{4}\int_{-z_0}^{z_0} \frac{1-\psi_0^2}{g^2(z)}  \, \mathd z \, ,~~~~~~H_1=-\frac{1}{8}\int_{-z_0}^{z_0} \frac{1+\psi_0^2}{g^2(z)}   \, \mathd z \, .
\nonumber
\end{eqnarray}

The higher order terms in the $\chi$PT lagrangian are generated by integrating out the resonances. For the derivation of the ${\cal O}(p^6)$ terms, only those operators containing one resonance field are involved:
\begin{eqnarray}
S_{\rm{YM}}\bigg|_{{\rm 1-res.}}&=&\sum_n \int \mathd ^4x
\, \bigg\{\,
-\,  \bra \frac{f^{\mu\nu}_+}{2}
\bigg[    (\nabla_\mu v_\nu^n-\nabla_\nu v_\mu^n)a_{Vv^n}-\frac{i}{2}([u_\mu,a_\nu^n]
\, -\, [u_\nu,a_\mu^n])a_{Aa^n}  \bigg] \ket
\nn\\
&&
\qquad  -\, \bra
\frac{i}{4}[u^\mu,u^\nu]
\bigg[    (\nabla_\mu v_\nu^n-\nabla_\nu v_\mu^n)b_{v^n\pi\pi}
\, -\, \frac{i}{2}([u_\mu,a_\nu^n]-[u_\nu,a_\mu^n])b_{a^n\pi^3}]\ket
\nn\\
&&    +   \bra  \frac{f^{\mu\nu}_-}{2}
\bigg[  (\nabla_\mu a_\nu^n-\nabla_\nu a_\mu^n)a_{Aa^n}
\, -\, \frac{i}{2}([u_\mu,v_\nu^n]-[u_\nu,v_\mu^n])(a_{Vv^n}-b_{v^n\pi\pi})\bigg]\ket
\,\bigg\}\, ,
\\
S_{\rm{CS}}\bigg|_{{\rm 1-res.}}&=&
\sum_n \int \mathd ^4x
\bigg\{
-\frac{N_C}{32\pi^2}\, c_{v^n}\epsilon^{\mu\nu\alpha\beta}
\bra u_\mu\{v_\nu^n,f_{+\alpha\beta}\}\ket
+ \frac{N_C}{64\pi^2}\, c_{a^n}\epsilon^{\mu\nu\alpha\beta}
\bra u_\mu\{a_\nu^n,f_{-\alpha\beta}\}\ket
\nn \\
&&\qquad \qquad \qquad +\frac{i N_C}{16\pi^2}\, (c_{v^n}-d_{v^n})\epsilon^{\mu\nu\alpha\beta}
\bra v_\mu^n u_\nu u_\alpha u_\beta\ket \,\bigg\} \, ,
\label{eq.S-1res-odd}
\end{eqnarray}
where all the resonance couplings are given by integrals of the 5D wave functions $\psi_0$ and $\psi_n$. Through the equation of motion for $\psi_n$,
all the couplings in the odd sector can be related to those in the even sector:
\begin{eqnarray}
c_{v^n}&=&\frac{m_{v^n}^2}{2f_\pi^2 }b_{v^n\pi\pi}\, , \qquad\qquad
c_{a^n}=\frac{m_{a^n}^2}{3f_\pi^2 }b_{a^n\pi^3}\, ,\nn\\
d_{v^n}&=&\frac{m_{v^n}^2}{12f_\pi^2 }\int_{-z_0}^{+z_0} \frac{\psi_{2n-1}(1-\psi_0^4)}{g^2(z)} \mathd z \label{eq:dv}.
\end{eqnarray}
When the resonance fields are integrated out, the ${\cal O}(p^6)$ LECs are obtained and expressed through the summation of the resonance couplings.
Those diagrams with one odd and one even vertices give rise to the odd couplings, e.g.,
\begin{equation}
C_{22}^W=\frac{N_C}{64\pi^2}S_{\pi VV},~~S_{\pi VV}=\sum_{n=1}^\infty \frac{a_{Vv^n}c_{v^n}}{m_{v^n}^2}.
\end{equation}
By employing the relation for $c_{v^n}$ in eq.~(\ref{eq:dv}) the involved summation can be further simplified leading to
\begin{equation}
C_{22}^W=\frac{N_C}{32\pi^2f_\pi^2}L_9.
\end{equation}
For the LECs in the even sector, two different diagrams may contribute, namely the even-even diagrams or odd-odd diagrams, e.g.,
\begin{equation}
C_{52}=-\frac{1}{8}S_{V\pi\pi}- 2 ~(\frac{N_c}{32 \pi^2})^2~ S_{V\pi^4},
\end{equation}
where
\begin{equation}
 S_{V\pi\pi}=\sum_{n=1}^\infty\frac{a_{Vv^n}b_{v^n\pi\pi}}{m_{v^n}^2},~~S_{V\pi^4}=\sum_{n=1}^\infty \frac{c_{v^n} (c_{v^n}-d_{v^n})}{m_{v^n}^2}.
 \end{equation}
 Again the summation involving odd couplings can be further simplified through eq.~(\ref{eq:dv}) resulting
\begin{equation}
 C_{52}  = -\frac{1}{8}S_{V\pi\pi}-\frac{N_C^2}{1920\pi^4f_\pi^2},
\end{equation}
while the calculation of $S_{V\pi\pi}$ requires the Green function at zero momentum. The expressions for the full set of
LECs can be found in our original paper~\cite{Colangelo:2012ip}. Since the theory is defined in the large $N_C$ and chiral limit, the couplings for the multi-trace operators and those involving scalar/pseudoscalar sources are vanishing.

\section{Model-independent relations among LECs}
Already at ${\cal O}(p^4)$,  some relations exist among the LECs which are independent of the background metric of the models. The relations among $L_1$, $L_2$ and $L_3$ follow from large $N_C$ analysis and the fact that only vector resonances are introduced~\cite{Guo:2007hm}. The same arguments lead to the relation for the ${\cal O}(p^6)$ LECs relevant for $\pi\pi$ scattering
 \begin{equation}
C_1 + 4 C_3 = 3 C_3 + C_4.
\end{equation}
The relation between $L_9$ and $L_{10}$ is related to the axial form factor of the $\pi \to l\nu \gamma$ decay, which
is vanishing identically in this class of models. At order $p^6$, this leads to the relation
\begin{equation}
2 C_{78} - 4 C_{87} + C_{88}  = 0.
\end{equation}
Some other relations are found for the $\gamma\gamma\to \pi \pi $ scattering process, where certain combinations of LECs are involved, namely $a^{00}_2$,
$b^{00}$, $a^{+-}_2$ and $b^{+-}$~\cite{Colangelo:2012ip}.
In the class of models considered, it turns out that only the odd-odd diagrams contribute to these combinations which are then completely fixed
\begin{eqnarray}
&&a^{00}_2=N_C^2\,,\,~~~~~~~~b^{00}=\frac{N_C^2}{6}\,,\nn\\
&&a^{+-}_2=0\,,\,~~~~~~~~~~b^{+-}=0.
\end{eqnarray}
More relations for the LECs within the even sector are still under investigation.
For the LECs in the odd sector, as shown before, one finds
\begin{equation}
C_{22}^W=\frac{N_C}{32\pi^2f_\pi^2}L_9,\label{eq:C22}
\end{equation}
which is shown to be the long-distance manifestation of the relation between the $\gamma^*\to\pi\pi$ and
$\pi\to\gamma\gamma^*$ form factors~\cite{Colangelo:2012ip}. Taken into account that $L_9=-L_{10}$, this is also the infrared limit of the
Son-Yamamoto relation between the transverse triangle structure function and the left-right correlator~\cite{Son:2010vc}.
One can further show that all the ${\cal O}(p^6)$ LECs in the odd sector, if not vanishing in the large-$N_c$ and chiral limit,
can be expressed through $L_1$, $L_9$ and another constant
\begin{equation}
Z=\frac{1}{4}\sum_{n=1}^\infty b_{a^n\pi^3}^2=\int_{-z_0}^{+z_0} \frac{\psi_0^2(1-\psi_0^2)^2}{4g^2(z)}\mathd z.
\end{equation}
For example, one finds
\begin{equation}
C_{23}^W=\frac{N_C}{96\pi^2f_\pi^2}(L_9-8L_1),
\label{eq:C23}
\end{equation}
which is a manifestation of the relation between certain form factors involving the axial source~\cite{Colangelo:2012ip}. The constant $Z$ is present only when the corresponding
operators involve at least five fields, namely $O_{12}^W$, $O_{16}^W$ and $O_{17}^W$, and is
supposed to originate from some higher order terms, e.g., $ {\cal F}^3$.

\newpage

\section*{Acknowledgements}

{\small
Contribution to the poster section of
QCD@Work 2012 --International Workshop on QCD: Theory and Experiment--,
18-21 June 2012, Lecce (Italy).
I would like to thank the organizers for their attentions
during the workshop.
This work is partially supported by the Italian MIUR PRIN 2009, and by the National Natural Science Foundation of China under Grant No. 11135011.
}




\bibliographystyle{aipproc}   





\end{document}